\documentclass[sigconf]{acmart}
\AtBeginDocument{%
  \providecommand\BibTeX{{%
    \normalfont B\kern-0.5em{\scshape i\kern-0.25em b}\kern-0.8em\TeX}}}

\copyrightyear{2022}
\acmYear{2022}
\setcopyright{rightsretained}
\acmConference[NLBSE'22]{The 1st Intl. Workshop on Natural Language-based Software Engineering}{May 21, 2022}{Pittsburgh, PA, USA}
\acmBooktitle{The 1st Intl. Workshop on Natural Language-based Software Engineering (NLBSE'22), May 21, 2022, Pittsburgh, PA, USA}
\acmDOI{10.1145/3528588.3528665}
\acmISBN{978-1-4503-9343-0/22/05}

\begin{document}
\title{On the Evaluation of NLP-based Models for Software Engineering}

\author{Maliheh Izadi}
\email{m.izadi@tudelft.nl}
\affiliation{ 
\institution{Delft University of Technology}
\country{Delft, Netherlands}
}

\author{Matin Nili Ahmadabadi}
\email{matin_nili@alumni.ut.ac.ir}
\affiliation{ 
\institution{University of Tehran}
\country{Tehran, Iran}
}
\renewcommand{\shortauthors}{Izadi and Nili}

\begin{abstract}
NLP-based models have been increasingly incorporated to address SE problems. These models are either employed in the SE domain with little to no change, or they are greatly tailored to source code and its unique characteristics. Many of these approaches are considered to be outperforming or complementing existing solutions. However, an important question arises here: \textit{Are these models evaluated fairly and consistently in the SE community?}. To answer this question, we reviewed how NLP-based models for SE problems are being evaluated by researchers. The findings indicate that currently there is no consistent and widely-accepted protocol for the evaluation of these models. While different aspects of the same task are being assessed in different studies, metrics are defined based on custom choices, rather than a system, and finally, answers are collected and interpreted case by case. Consequently, there is a dire need to provide a methodological way of evaluating NLP-based models to have a consistent assessment and preserve the possibility of fair and efficient comparison.
\end{abstract}

\keywords{Evaluation, Natural Language Processing, Software Engineering}

\maketitle

\section{Introduction}
Researchers have been using NLP-models 
to solve a diverse set of SE problems
such as code generation, completion, summarization, 
bug fixing, question answering, test case generation, 
documentation, and many more.
As these models attract more researchers 
and the number and diversity of studies grows, 
it is imperative to have good evaluation measures and techniques 
to assess them properly.
These measures should be consistent throughout the literature 
in order to conduct fair and comparable comparisons.
To understand the evaluation of NLP models,
we reviewed the field in the past five years and report the results here.
To the best of our knowledge,
we are the first to conduct a systematic literature review 
on evaluation of NLP-based models 
to understand the underlying patterns, 
identify the challenges, 
and recommend future research direction.

\section{Methodology}
We conducted our systematic review using the following protocol.
Our main research question is 
``How are NLP-based models evaluated in SE?''.
Search phrases in the title, abstract or body of a paper are 
\textit{NLP}, \textit{natural language processing}, \textit{code}, and \textit{evaluation}.
Papers must be peer-reviewed, written in English, 
and be published after 2017
by one of the following SE prominent conferences and journals: 
\textit{ICSE}, \textit{ESEC/FSE}, \textit{ASE},
\textit{IEEE TSE}, \textit{ACM TOSEM}, and \textit{EMSE}.
We used \textit{Google Scholar} as the source, 
and retrieved $157$ papers.
Two of the authors manually inspected all papers 
to identify the papers that propose 
an NLP-based model to solve a SE problem.
Finally, $53$ papers were excluded
because of one or more of the following reasons: 
the paper's scope was unrelated to NLP and SE,
the main proposed model was not based on NLP, 
or it was a secondary or duplicate study.
Next, we present the result of the review 
on the remaining $104$ included papers.
More information on the protocol and papers can be found in our GitHub repository.\footnote{\url{https://github.com/MalihehIzadi/nlp4se_eval}}

\section{Evaluation of NLP-based Models}
There are two approaches to evaluation
\textit{intrinsic} with a focus on intermediary goals (sub-tasks),
and \textit{extrinsic} for assessing the performance of the final goal.
NLP-based models in SE are generally evaluated 
with one or more of the following metrics.\\
(1) \textbf{Automatically}:
Automatic evaluation consists of three groups, namely 
(i) metrics for assessing the results of classification models 
such as \textit{Accuracy}, \textit{Precision}, \textit{Recall}, and \textit{F} measure, 
(ii) metrics for assessing recommendation lists 
including \textit{$Top@n$} or ranked versions such as \textit{MRR} and \textit{MAP}, 
and 
(iii) metrics for analyzing the quality of generated text or source code
including \textit{BLEU}, \textit{METEOR}, \textit{ROUGE}, 
\textit{CIDEr}, \textit{chrF}, 
\textit{Perplexity}, and \textit{Levenshtein similarity} metrics.\\
(2) \textbf{Manually}:
Manual assessment is more subjective and
heeds the judgment of human participants.
Researchers first select the relevant metric(s) 
to evaluate different aspects of the proposed model's output.
Then, they invite a group of experts 
to assess the results based on the selected metrics.
For instance, for a code summarization task, 
researchers use \textit{informativeness} 
as an indicator of the quality of the generated summaries 
from the developers' perspective.

Automatic evaluation is easier, faster, and completely objective 
compared to the manual version.
Thus most researchers opt to use automatic evaluation 
for assessing their models.
However, human-based assessments 
can potentially convey more information for several aspects of a model,
hence, they can be used to complement automatic evaluation.
Recently, Roy et al.~\cite{roy2021reassessing}
conducted an empirical study on the applicability and interpretation of 
automatic metrics for evaluation of the. code summarization task.
With the help of $226$ human annotators,
they assessed the degree to which automatic metrics reflect human evaluation. 
They claim that less than $2$ points improvements for an automatic metric such as BLEU
do not guarantee systematic improvements in summarization quality. 
This makes the role of human assessment salient.

Although automatic measures are uniformly defined in the literature,
manual metrics are harder to define, interpret and use.
These measures must be properly indicative of a model's goal and performance.
Furthermore, their definitions and usage must be kept consistent 
to have comparable results. 
Hence, in the following we review 
the most popular existing manual assessment measures in the SE domain
and leave the rest of them
(such as effectiveness, comprehensibility, time-saving, 
relatedness, rightness, usability, recency, 
grammatically correctness,  advantageousness, diversity, self-explanatory, theme identification, and more)
for a more comprehensive study.
\textbf{Usefulness}: 
    Several studies define usefulness as 
    how useful participants find the proposed solution for solving the problem at hand~\cite{wang2021automatic,ren2020api,di2017surf,cai2019answerbot,izadi2021topic}.
    Others define usefulness 
    as the tendency or preference of users 
to use their proposed model~\cite{zhao2019recdroid}.
    Jiang et al.~\cite{jiang2019machine} 
    assess the usefulness of its results based on 
    both its accuracy and the difficulty of generating outputs. 
    That is, they focus on how often the model works when it is indeed needed.
\textbf{Naturalness}, \textbf{Expressiveness}, \textbf{Readability}, and \textbf{Understandability}:
    Roy et al.~\cite{roy2020deeptc} 
    define naturalness as 
    how easy it is to read and understand generated outputs.
    They also use readability to measure 
    to what extent the output is perceived 
as readable and understandable by the participants.
    Aghamohammadi et al.~\cite{aghamohammadi2020generating} 
    define naturalness as
    how smooth, human-readable, and syntactically-correct are their outputs. 
    Gao et al.~\cite{gao2020generating} 
    measure naturalness as the grammatical correctness and fluency of a generated sentence.
    Zhou et al.~\cite{zhou2019drone} use expressiveness 
    as whether their model's output is clear and understandable.
\textbf{Correctness} or \textbf{Content}: 
    Huang et al.~\cite{huang2018api} define correctness as 
    whether participants can find the correct API using their proposed tool, 
    while  Chen et al.~\cite{chen2017unsupervised} define it 
    as a measure to verify the general correctness of the abbreviations and synonyms in their thesaurus.
    In Roy et al.'s~\cite{roy2020deeptc} study,
    content means whether a summary correctly reflects the content of a test case.
\textbf{Completeness} and \textbf{Informativeness}:
   Uddin et al.~\cite{uddin2017automatic} 
    define completeness 
    as a complete yet presentable summarization of API reviews.
    Aghamohammadi et al.~\cite{aghamohammadi2020generating} 
    define informativeness as 
    how much of the important parts of a piece of code are covered by a generated summary. 
\textbf{Conciseness}: 
    In Roy et al.'s~\cite{roy2020deeptc} study,
    concise summaries do not include extraneous or irrelevant information. 
    Zhou et al.\cite{zhou2019drone}
    quantifies conciseness through answering 
    whether the repair recommendation is free of other constraint-irrelevant information.
\textbf{Relevance} or \textbf{Similarity}:
    Several studies define relevance as to how relevant 
    is the model's output to the reference text or code~\cite{gao2020generating,wang2021automatic,ren2020api,cai2019answerbot}.
    Others asked developers to rate the similarity, relatedness, 
    and contextual or semantic similarity
    between outputs and reference texts~\cite{liu2019automatic,wainakh2021idbench,jiang2017automatically}.

\section{Discussion, and Future Direction}
We reviewed $104$ studies to understand
how NLP-based models are usually evaluated in the SE domain
and provided the list of most used metrics.
Next, we provide the main challenges for the evaluation of NLP-based models.
(1) Both automatic and manual approaches can be utilized
to provide a more holistic view of the performance of an NLP-based model, 
however, not all of the eligible studies use both of these approaches.
(2) For the manual form of assessment, 
there exist numerous and sometimes conflicting or ambiguous evaluation metrics.
This problem exacerbates in the case of 
measures with multiple definitions (e.g., informativeness)
or in case of multiple metrics which are overlapping 
(e.g., naturalness, readability, understandability, and expressiveness).
Some researchers evaluate different aspects of their model, 
(e.g., completeness, naturalness, or correctness) 
while others only address one or two aspects.
As there are various aspects to each model,
there should be a methodological way 
to first identify the most important aspects of a given SE task
and then properly evaluate those aspects with concretely defined metrics.
(3) In addition to the definition and use of the metrics,
there is no standard for defining the set of answers.
That is, some use yes/no answers, while others use $n$-point Likert scale 
or even free-format text answers.
(4) Finally, identifying novel evaluation metrics or techniques 
can help SE researchers assess these models more thoroughly and where it matters.
For example, for the automatic code completion task, 
predicting an identifier is more valuable and difficult than predicting a keyword.
Hence, evaluating models for predicting any token is not very helpful. 
Through reviewing the literature, 
we took the first step 
toward addressing the challenges of proper evaluation 
for NLP-based models.
We suggest future research focus on providing 
a systematic and consistent framework for evaluation of these models to 
(1) clearly define measures, 
(2) distinguish between different needs of SE tasks, 
and (3) determine the proper use of a measure in the context.
Hopefully, a systematic way of evaluation 
makes it possible to conduct fair and correct evaluations 
for the NLP-based models in the SE field.

\bibliographystyle{ACM-Reference-Format}
\bibliography{main}
\end{document}